\documentclass[12pt,english]{article}
\usepackage[T1]{fontenc}
\usepackage[latin9]{inputenc}
\usepackage{color}
\usepackage{multirow}
\usepackage{float}
\usepackage{mathrsfs}
\usepackage{amsthm}
\usepackage{amsmath}
\usepackage{amssymb}
\usepackage{graphicx}
\usepackage{setspace}
\usepackage{esint,chicago}
\usepackage{amsmath,epsfig,amssymb,amsfonts,amsthm,epsfig,verbatim}
\usepackage{color,graphicx,lscape,longtable}

\newcommand{\cbold}    {\mbox{\bf c}}

\newcommand{\Abold}    {\mbox{\bf A}}

\newcommand{\Ibold}    {\mbox{\bf I}}

\newcommand{\Lbold}    {\mbox{\bf L}}

\newcommand{\Ybold}    {\mbox{\bf Y}}

\newcommand{\zerobold}  {\mbox{\bf 0}}

\newcommand{\cboldhat}    {\widehat{\mbox{\bf c}}}

\newcommand{\mubold}      {\mbox{\boldmath${\mu}$}}

\newcommand{\phibold}     {\mbox{\boldmath${\phi}$}}

\newcommand{\Psibold}     {\mbox{\boldmath${\Psi}$}}

\newcommand{\Sigmabold}   {\mbox{\boldmath${\Sigma}$}}
\newcommand{\Gammabold}   {\mbox{\boldmath${\Gamma}$}}

\newcommand{\Sigmaboldhat} {\widehat{\mbox{\boldmath${\Sigma}$}}}
\newcommand{\Lboldhat} {\widehat{\mbox{\boldmath${L}$}}}

\newcommand{\Lvec}{\mbox{\bf $\ell$}}

\usepackage{ifpdf} 
\ifpdf 

 \IfFileExists{lmodern.sty}{\usepackage{lmodern}}{}

\fi 

\usepackage{mathrsfs}
\usepackage[textwidth=6.5in,textheight=9in]{geometry}

\usepackage{ifthen}

\def\LyX{\texorpdfstring{%
  L\kern-.1667em\lower.25em\hbox{Y}\kern-.125emX\@}
  {LyX}}

\makeatother

\usepackage{babel}
  \addto\captionsbritish{}
  \addto\captionsbritish{}
  \addto\captionsbritish{}
  \addto\captionsenglish{}
  \addto\captionsenglish{}

\addto\captionsbritish{}

\usepackage{enumitem}
\setenumerate{label= \textup{(}\textit{\roman*}\textup{)},leftmargin=1.6pc,topsep=5pt}

\usepackage[left,displaymath,mathlines,pagewise]{lineno}
\usepackage{xcolor}

\begin{document}

\begin{center}
\large{\bf{Functional Mapping of Multiple Dynamic Traits}}\\
\vspace{0.5cm}
{Jiguo Cao$^{1}$, Liangliang Wang$^{1}$, Zhongwen Huang$^{2,3}$, \\Junyi Gai$^3$, Rongling Wu$^4$\\$^1$Department of Statistics and Actuarial Science, Simon Fraser University,\\Burnaby, BC, V6T1N8, Canada. Email: jca76@sfu.ca
\\$^2$Department of Agronomy, Henan Institute of Science and Technology, Xinxiang 453003, China.\\$^3$Soybean Research Institute of Nanjing Agricultural University,\\National Key Laboratory for Crop Genetics and Germplasm Enhancement, Nanjing 210095, China\\$^4$Center for Statistical Genetics, Pennsylvania State University\\Hershey, PA 17033, USA. Email: rwu@hes.hmc.psu.edu\\}

\vspace{0.5cm}

\begin{minipage}[t]{\columnwidth}%

\paragraph*{{ Abstract}}

\normalsize{ Many biological phenomena undergo developmental changes in time and space. Functional mapping, which is aimed at mapping genes that affect developmental patterns, is instrumental for studying the genetic architecture of biological changes. Often biological processes are mediated by a network of developmental and physiological components and, therefore, are better described by multiple phenotypes. In this article, we develop a multivariate model for functional mapping that can detect and characterize quantitative trait loci (QTLs) that simultaneously control multiple dynamic traits. Because the true genotypes of QTLs are unknown, the measurements for the multiple dynamic traits are modeled using a mixture distribution. The functional means of the multiple dynamic traits are estimated using the nonparametric regression method, which avoids any parametric assumption on the functional means. We propose the profile likelihood method to estimate the mixture model. A likelihood ratio test is exploited to test for the existence of pleiotropic effects on distinct but developmentally correlated traits. A simulation study is implemented to illustrate the finite sample performance of our proposed method. We also demonstrate our method by identifying QTLs that simultaneously control three dynamic traits of soybeans. The three dynamic traits are the time-course biomass of the leaf, the stem, and the root of the whole soybean. The genetic linkage map is constructed with 950 microsatellite markers. The new model can aid in our comprehension of the genetic control mechanisms of complex dynamic traits over time.}%
\vspace{0.5cm}

\end{minipage}
\par\end{center}

\newpage
\section{Introduction}The past two decades have witnessed the rapid development of genomic technologies that allow for molecular characterization of polymorphic markers throughout the entire genome. Nowadays molecular markers are readily available for a diversity of species. These markers are used to identify and localize quantitative trait loci (QTLs) that control phenotypic variation in a complex trait of interest. For a variety of traits hundreds of thousands of QTLs have been discovered, which play an important role in explaining the genetic control of biological characteristics. Among many successful examples of genetic mapping, it has been observed that QTLs are responsible for: branching, florescence, and grain architecture in maize \shortcite{Doebley97,Gallavotti04,Wang05}; fruit size and shape in tomatoes \shortcite{Paterson88,Frary00}; the reduction of grain shattering \shortcite{li06}; complex behaviors in Drosophila \shortcite{Anholt04}; whole blood serotonin level in humans \shortcite{Weiss05}; as well as height growth and 16 other quantitative traits in the Hutterites, a founder human population \shortcite{Weiss06}.

Genetic mapping of QTLs using molecular markers is founded on a statistical model pioneered by \shortciteN{Weller86} and \shortciteN{Lander89}. This model implements the EM algorithm to estimate the chromosomal positions and genetic effects of individual QTLs on a phenotypic trait. The publication of the first mapping model has led to an explosion of new statistical methods, which can precisely and accurately map QTLs under a variety of circumstances \shortcite{Knapp91,Haley92,Jansen94,Zeng94,Sen01,Kao02}. By considering the developmental complexity of several complex traits, such as growth, cell cycles and drug response, \shortciteN{Ma02} developed a new statistical method called functional mapping. This method maps that QTLs that influence the dynamic behavior of phenotypic values in time and space. Functional mapping capitalizes on mathematical aspects of biological and biochemical principles to model the temporal-spatial pattern of genetic effects triggered by QTLs. It has been proven to be a powerful method for studying and mapping the genetic architecture of dynamic trajectories across time and space, and for testing the genetic mechanisms underlying developmental alterations \shortcite{WuLin06,LiWu10,He10}. Statistically, functional mapping displays an increased power to detect QTLs, because fewer parameters are used to describe dynamic traits.

To broaden the range of applications for functional mapping, in which no parametric forms are available to specify the dynamic behavior of a trait, several nonparametric versions of functional mapping have been proposed, such as Legendre orthogonal polynomials \shortcite{Lin06,Yang07,das2011dynamic} and B-splines \shortcite{Yang09}. These nonparametric functions are flexible enough to represent dynamic or longitudinal traits in various shapes. The nonparametric functions are estimated directly from repeated measurements of dynamic traits, thus avoiding biases arising from inaccurate parametric assumptions.

With an increasing interest in systems mapping, which aims to elucidate a comprehensive picture of trait development, some studies have started to map phenotypic changes of multiple traits over time and space \shortcite{Zhao05,Lietal06,Wu11}. \shortciteN{Zhao05} developed a growth equation approach for mapping two correlated growth traits. In a recent study, \shortciteN{Wu11} implemented a system of differential equations to model the temporal change of QTL effects on multiple traits that constitute a dynamic system. However, since these approaches require explicit mathematical equations to specify the dynamic traits, they are not as well suited to applications where no explicit equations exist.

The purpose of this article is to develop a flexible functional mapping method that can detect QTLs responsible for multiple dynamic traits. Each dynamic trait is represented by a nonparametric function, which is expressed as a linear combination of basis functions. We propose the profile likelihood method to estimate the nonparametric functions, as well as the correlations among multiple dynamic traits. A likelihood ratio test is implemented to identify QTLs at a grid of possible QTL locations. The significance threshold of the likelihood ratio test is derived using the permutation test. Since the exact genotype of a potential QTL at a grid point is unknown, dynamic traits are assumed to be in a mixture normal distribution. The Cholesky decomposition \shortcite{Trefethen97} is used to parameterize the variance-covariance matrix of the multiple dynamic traits to ensure that the estimated variance-covariance matrix is both symmetric and positive definite.

The remainder of this article is organized as follows. Section 2 introduces our statistical model for functional mapping. The parameter estimation method for our statistical model is introduced in Section 3. In Section 4, our functional mapping method is applied to detect QTLs that control multiple dynamic traits of soybeans. Section 5 presents simulation studies implemented to evaluate the finite sample performance of our functional mapping method. The discussion of this model is given in Section 6.

\section{A Mixture Model}Suppose multiple dynamic traits are measured at a series of time points. Let $Y_{hi}(t_{ir})$ be the measured $h$-th dynamic trait at the $r$-th time point
$t_{ir}$ for the $i$-th subject, $h=1,\cdots,H$, $i=1,\cdots,n_h$, $r=1,\cdots,m_{i}$. Let $\Ybold_i(t)=(Y_{1i}(t),\cdots,Y_{Hi}(t))^T$ denote the vector of measurements for multiple dynamic traits at the time point $t$\,. Given that the $i$-th subject has QTL genotype $j$, $j = 1,\cdots,J$, $\Ybold_i(t)$ is assumed to have a multivariate normal distribution with mean $\mubold_j(t) = (\mu_{1j}(t),\cdots,\mu_{Hj}(t))^T$ and a variance-covariance matrix $\Sigmabold$.
 
In practice, the true genotypes of QTLs are unknown. But we can calculate the conditional probabilities of QTL genotypes given marker genotypes as a function of the recombination fractions between the QTL and markers \shortcite{Wu07}. Let $\omega_{ij}$ be the line origin probability of the $i$-th subject having the QTL genotype $j$, $j = 1,\cdots,J$. The line origin probability $\omega_{ij}$ can be calculated in advance based on experimental population designs such as inbreed, outbreed and backcross. The vector of measurements for multiple dynamic traits, $\Ybold_i(t)$, is modeled using a mixture distribution:
\begin{eqnarray}\label{Eqn:MixDistribution}\Ybold_i(t) \sim \sum_{j=1}^J \omega_{ij}f(\Ybold_i(t)|\mubold_j(t),\Sigmabold)\,,\end{eqnarray}
where $f(\Ybold_i(t)|\mubold_j(t),\Sigmabold)$ is the probability density function (pdf) of the multivariate normal distribution with mean $\mubold_j(t)$ and a variance-covariance matrix $\Sigmabold$, which is expressed as follows:
\begin{eqnarray}\label{eqn:multiNormalpdf}
f(\Ybold_{i}(t)|\mubold_j(t),\Sigmabold) &=& (2\pi)^{-H/2}|\Sigmabold|^{-1/2}\exp\{-(\Ybold_i(t)-\mubold_j(t))^T\Sigmabold^{-1}(\Ybold_i(t)-\mubold_j(t))/2\}.
\end{eqnarray}

In order to avoid any parametric constraints on the functional mean of the $h$-th dynamic trait $\mu_{hj}(t)$, given the QTL genotype $j$, $\mu_{hj}(t)$ is estimated using the nonparametric smoothing method \shortcite{RamsaySilverman05}. In our article, the functional mean $\mu_{hj}(t)$ is represented as a linear combination of basis functions,
\begin{eqnarray*}\mu_{hj}(t) = \sum_{k=1}^K c_{hjk} \phi_{hjk}(t) = \cbold_{hj}^T\phibold_{hj}(t)\,,\end{eqnarray*}
where $\phibold_{hj}(t) = (\phi_{hj1}(t),\cdots,\phi_{hjK}(t))^T$ is a vector of basis functions, and $\cbold_{hj} = (c_{hj1},\cdots,c_{hjK})^T$ is a vector of basis coefficients. Cubic B-splines are often chosen as
basis functions since any B-spline basis function is only positive over a short interval and zero elsewhere. This is called the \emph{compact support} property, and is essential for efficient computation \shortcite{deboor01}.

The variance-covariance matrix, $\Sigmabold$, must be symmetric and positive-definite. Therefore it may be estimated using a constrained optimization method. Alternatively, $\Sigmabold$ can be decomposed as
\begin{eqnarray*}\label{Dfactor}  \Sigmabold^{-1} = \Lbold \Lbold^T \,,\end{eqnarray*}
where $\Lbold$ is a lower triangular matrix with strictly positive diagonal entries\,. This is called the Cholesky decomposition \shortcite{Trefethen97}. By employing the Cholesky decomposition, we can estimate the lower triangular matrix $\Lbold$ directly, without considering the usual constraints on $\Sigmabold$, since $\Sigmaboldhat = (\Lboldhat \Lboldhat^T)^{-1}$ will automatically be symmetric and positive-definite \cite{cao2012linear}. Essentially the Cholesky decomposition allows us to convert a constrained optimization problem to an unconstrained optimization. The Cholesky decomposition is not uniquely defined for a given positive-definite matrix, but it can be made unique by requiring the diagonal elements in $\Lbold$ to be all positive. Consequently the diagonal elements in $\Lbold$ are parameterized in terms of their logarithms in our article.

\section{QTL Mapping Method for Multiple Dynamic Traits}
The mixture model (\ref{Eqn:MixDistribution}) has two types of parameters to estimate, the basis coefficient, $\cbold_{hj}$, and the lower triangular matrix, $\Lbold$. Define the long vector of basis coefficients $\cbold_j = (\cbold_{1j}^T,\cdots,\cbold_{Hj}^T)^T$, and $\cbold = (\cbold_{1}^T,\ldots,\cbold_{J}^T)^T$. We propose to estimate the basis coefficient, $\cbold$, and the lower triangular matrix, $\Lbold$, using the profile likelihood method. The method estimates the two parameters in two nested levels of optimization. In the inner level of optimization, the basis coefficient, $\cbold$, is estimated by maximizing the log likelihood function for a given lower triangular matrix, $\Lbold$. There is no analytic expression for the estimated basis coefficient, $\cboldhat$, but it can be viewed as an implicit function of $\Lbold$. In the outer level of optimization, the lower triangular matrix, $\Lbold$, is estimated by maximizing the profile likelihood function, in which the basis coefficient is removed from the parameter space by treating it as a function of $\Lbold$. Although the estimated basis coefficient, $\cboldhat$, has no analytic formula, we use the implicit function theorem to obtain analytic gradients for the optimization iteration process, which makes computation faster and more stable. We outline the details of the profile likelihood method below.

\subsection{Profile Likelihood Method for Estimating the Mixture Model}
To simply notation, we first define some vectors and matrices to be used in the likelihood function. These matrix representations also help to significantly increase computational efficiency in MATLAB \shortcite{MATLAB2013}. Define the long vector of data $\Ybold_i = (\Ybold_i(t_{i1})^T,\cdots,\Ybold_i(t_{im_i})^T)^T$, and the long vector of basis coefficients $\cbold_j = (\cbold_{1j}^T,\cdots,\cbold_{Hj}^T)^T$. Then the distribution of $\Ybold_i$, given the QTL genotype $j$, is the multivariate normal distribution with mean $\Psibold_{ij}\cbold_j$ and a variance-covariance matrix $\Gammabold$\,, where $\Gammabold = \Ibold_{m_i} \otimes\Sigmabold$, $\Psibold_{ij}$ is a $Hm_i\times HK$ matrix, and $\Abold_{ijr}$, $r=1,\ldots,m_i$, is a block diagonal matrix with the $h$-th diagonal block as $\phibold_{hj}^T(t_{ir})$ defined below:
\begin{displaymath}
\Psibold_{ij} =
\left( \begin{array}{cccc}
\Abold_{ij1}\\
\Abold_{ij2}\\
\vdots\\
\Abold_{ijm_i}\\
\end{array} \right),
\Abold_{ijr} =
\left( \begin{array}{cccc}
\phibold_{1j}^T(t_{ir}) & \zerobold_{1\times K} & \ldots &\zerobold_{1\times K}\\
\zerobold_{1\times K} & \phibold_{2j}^T(t_{ir}) & \ldots & \zerobold_{1\times K}\\
\vdots & \vdots & \ddots & \vdots\\
\zerobold_{1\times K} & \zerobold_{1\times K} & \ldots & \phibold_{Hj}^T(t_{ir})
\end{array} \right)
\end{displaymath}

The vector of basis coefficients $\cbold$ is estimated by maximizing the log likelihood function for a given lower triangular matrix, $\Lbold$:
\begin{eqnarray}\label{Eqn:InnerCriterion}J(\cbold|\Lbold) = \sum_{i=1}^n\log\left[\sum_{j=1}^J \omega_{ij}f(\Ybold_{i}|\cbold_{j},\Lbold)\right]\,,\end{eqnarray}
where
\begin{eqnarray}f(\Ybold_{i}|\cbold_{j},\Lbold) &=& (2\pi)^{-H*m_i/2}|\Gammabold|^{-1/2}\exp\{-(\Ybold_i-\Psibold_i\cbold_{j})^T\Gammabold^{-1}(\Ybold_i-\Psibold_i\cbold_{j})/2\}\,,\label{eqn:f}\\
\Gammabold^{-1} &=&   \Ibold_{m_i} \otimes \Lbold \Lbold^T,\nonumber\\|\Gammabold| &=&|\Sigmabold|^{m_i} = |(\Lbold\Lbold)^{-1}|^{m_i} = |(\Lbold\Lbold)|^{-m_i} = |\Lbold|^{-2m_i}.\nonumber\end{eqnarray}

Since the log likelihood function (\ref{Eqn:InnerCriterion}) is structured as a mixture distribution, it is impossible to obtain an analytic formula for the estimate $\cboldhat$. We use the Newton-Raphson method to maximize $J(\cbold|\Lbold)$ as follows. Let $\cbold^{(0)}$ be the starting value for $\cbold$, the $v$-th iteration
step updates $\cbold$ by
\begin{eqnarray*}\cbold^{(v)} = \cbold^{(v-1)} - \bigg(\frac{d^2 J}{d\cbold^2}\bigg|_{\cbold^{(v-1)}}\bigg)^{-1}\bigg(\frac{dJ}{d \cbold}\bigg|_{\cbold^{(v-1)}}\bigg)\,,\end{eqnarray*}
$v=1,2,\ldots,$ until convergence occurs. To ensure the Newton-Raphson method is both stable and fast, the first derivative of $J(\cbold|\Lbold)$ with respect to $\cbold$ is derived analytically as follows.
\begin{eqnarray}\label{Eqn:dJdc}
\frac{dJ}{d\cbold_j} &=& \sum_{i=1}^n P_{ij}\Psibold_i^T\Gammabold^{-1}(\Ybold_i-\Psibold_i\cbold_{j}),
\end{eqnarray}
where
\begin{eqnarray*}
P_{ij} & = & \frac{\omega_{ij}f(\Ybold_i|\cbold_j,\Lbold)}{\sum_{j=1}^J\omega_{ij}f(\Ybold_i|\cbold_j,\Lbold)}\,,
\end{eqnarray*}
with $\sum_{j=1}^JP_{ij} = 1$\,.
The second derivative of $J(\cbold|\Lbold)$ with respect to $\cbold$ is hard to obtain analytically, hence we apply the finite-difference method to approximate the second derivative by using the analytic first derivative given in (\ref{Eqn:dJdc}).

The estimate for the basis coefficient, $\cboldhat$, is obtained for any given value of the lower triangular matrix, $\Lbold$, so $\cboldhat$ may be viewed as an implicit function of $\Lbold$, which is denoted as $\cboldhat(\Lbold)$. The lower triangular matrix $\Lbold$ is then estimated by maximizing the log profile likelihood function
\begin{eqnarray}\label{eqn:F}
F(\Lbold) = \sum_{i=1}^n\log\left[\sum_{j=1}^J \omega_{ij}f(\Ybold_{i}|\cboldhat_{j}(\Lbold),\Lbold)\right]\,.
\end{eqnarray}
by using the Newton-Raphson method. In the optimization process, the entries below or in the main diagonal of the $H\times H$ lower triangular matrix, $\Lbold$, are combined in a vector, $\Lvec$, with length $H(H+1)/2$. To ensure the Newton-Raphson method is both stable and fast, the first derivative of $F(\Lbold)$ is derived analytically using the chain rule after considering $\cboldhat$ as a function of $\Lvec$:
\begin{eqnarray*}
\frac{dF(\Lbold)}{d\Lvec} = \frac{\partial F(\Lbold)}{\partial \Lvec} + \frac{\partial F(\Lbold)}{\partial \cboldhat}\frac{d \cboldhat }{d \Lvec}\,.
\end{eqnarray*}
Since $\cboldhat$ is an implicit function of $\Lvec$, the derivative $d \cboldhat/d \Lvec$ can be derived analytically by applying the implicit function theorem as follows. We take advantage of the fact that the estimate $\cboldhat$ satisfies
\begin{eqnarray*}
\frac{\partial J} {\partial \cbold^T}\bigg|_{\cboldhat} \equiv 0
\end{eqnarray*}
Taking the $\Lvec$-derivative on both sides of the above identity, we obtain
\begin{eqnarray*}
\frac{d}{d\Lvec}\frac{\partial J} {\partial \cbold^T}\bigg|_{\cboldhat} = \bigg\{\frac{\partial^2 J} {\partial \cbold^T \partial \Lvec}\bigg|_{\cboldhat}\bigg\} +
\bigg\{\frac{\partial^2 J} {\partial \cbold^T \partial \cbold}\bigg|_{\cboldhat}\bigg\}\bigg\{\frac{d \cboldhat^T} {d \Lvec}\bigg\} \equiv 0,
\end{eqnarray*}
which yields
\begin{eqnarray*}
\frac{d \cboldhat^T} {d \Lvec} = -\bigg\{\frac{\partial^2 J} {\partial \cbold^T \partial \cbold}\bigg|_{\cboldhat}\bigg\}^{-1}\bigg\{\frac{\partial^2 J} {\partial \cbold^T \partial \Lvec}\bigg|_{\cboldhat}\bigg\},
\end{eqnarray*}
provided that $\partial^2 J/\partial \cbold^T \partial \cbold$ is non-singular at $\cbold = \cboldhat$.

The algorithm of our proposed profile likelihood method can be summarized as follows:

\noindent\rule{\textwidth}{0.3mm}
\hrule
\medskip
\noindent{\bf The algorithm of the profile likelihood method}
\medskip
\hrule
\medskip
\noindent 1. Choose an initial value, $\Lbold^{(0)}$, for the lower triangular matrix $\Lbold$.\\
2. For a given $\Lbold^{(\tau)}$, \\
2.1 \hspace{1cm} Estimate $\cbold$ by maximizing \\
{\hskip 15mm} $$J(\cbold|\Lbold) = \sum_{i=1}^n\log\left[\sum_{j=1}^J \omega_{ij}f(\Ybold_{i}|\cbold_{j},\Lbold)\right]$$
\hspace{1.7cm} using the Newton-Raphson method.\\
2.2 \hspace{1cm} After obtaining the estimate $\cboldhat$, calculate ${dF(\Lbold)}/{d\Lvec}$ and ${d^2F(\Lbold)}/{d\Lvec^2}$.\\
2.3 \hspace{1cm} Update $\Lvec$ by the Newton-Raphson method:
\begin{eqnarray*}\Lvec^{(
\tau)} = \Lvec^{(\tau-1)} - \bigg(\frac{d^2 F}{d\Lvec^2}\bigg|_{\Lvec^{(\tau-1)}}\bigg)^{-1}\bigg(\frac{dF}{d \Lvec}\bigg|_{\Lvec^{(\tau-1)}}\bigg)\,,
\end{eqnarray*}
3. $\tau = \tau+1$. Go to Step 2 until the Newton-Raphson iteration procedure for maximizing $F(\Lbold)$ converges.
\medskip
\hrule
\noindent\rule{\textwidth}{0.3mm}

\subsection{Likelihood Ratio Test}
For a given linkage map, we will search at any possible position in the genome for QTLs that simultaneously control multiple dynamic traits. The significance test for the existence of a QTL can be performed by formulating hypotheses as follows:
\begin{eqnarray*}H_0 &:& \cbold_1 = \cbold_2 = \cdots = \cbold_J, \\H_1 &:& \text{at least two of  } \cbold_j, j=1,\cdots,J, \text{ are not equal to each other}.\end{eqnarray*}
Under the alternative hypothesis, $H_1$, means of dynamic traits are different for at least two of $J$ QTL genotypes, i.e., at least two of $\mubold_j$, $j=1,\ldots,J$, are not equivalent. At any possible position of a QTL, we calculate the conditional probability, $\omega_{ij}$, of QTL genotypes given marker genotypes as a function of the recombination fractions between the QTL and markers \shortcite{Wu07}. The mixture model (\ref{Eqn:MixDistribution}) is then estimated with our proposed profile likelihood method.

Under the null hypothesis, $H_0$, means of dynamic traits are the same for different QTL genotypes, i.e. $\mubold_1(t) = \mubold_2(t) = \cdots = \mubold_J(t)$. Therefore, the vector of measurements for multiple dynamic traits, $\Ybold_i(t)$, is modelled by a multivariate normal distribution with mean, $\mubold^{H_0}(t)$, and a variance-covariance matrix, $\Sigmabold^{H_0}$:
\begin{eqnarray}\label{Eqn:H0}\Ybold_i(t) \sim f(\Ybold_i(t)|\mubold^{H_0}(t),\Sigmabold^{H_0})\,,
\end{eqnarray}
where $f(\cdot)$ is the probability density function (pdf) of multivariate normal distribution as expressed in (\ref{eqn:multiNormalpdf}). The mean, $\mubold^{H_0}(t)$, can be represented as a linear combination of basis functions, and the variance-covariance matrix, $\Sigmabold^{H_0}$, can be reparameterized using the Cholesky decomposition, as described in Section 2. Then $\mubold^{H_0}(t)$ and $\Sigmabold^{H_0}$ are estimated using the profile likelihood method as introduced in Section 3.

The likelihoods under the null and alternative hypotheses are calculated, from which the log-likelihood ratio (LR) is computed. Let $\widehat{\cbold}^{H0}$, $\widehat{\Lbold}^{H0}$, and $\widehat{\cbold}_j^{H1}$, $\widehat{\Lbold}^{H1}$ be the parameter estimates obtained under $H_0$ and $H_1$, respectively. The LR test statistic is given by
\begin{eqnarray}\label{Eqn:LR}
{\rm LR}=-2\left[\sum_{i=1}^n\log f(\Ybold_{i}|\widehat{\cbold}^{H0},\widehat{\Lbold}^{H0}) - \sum_{i=1}^n\log\left\{\sum_{j=1}^J \omega_{ij}f(\Ybold_{i}|\widehat{\cbold}_{j}^{H1},\widehat{\Lbold}^{H1})\right\}\right],
\end{eqnarray}
where $f(\cdot)$ is the probability density function (pdf) of $H\times m_i\,$-dimensional multivariate normal distribution as expressed in (\ref{eqn:f}). Given a significance threshold $T$, there is significant evidence that a QTL exists at a certain position if $\mathrm{LR}>T$. Since the distribution of the LR values under the null hypothesis is unknown, empirical permutation tests are usually used to determine the threshold (Churchill and Doerge 1994). In our article, we keep all multiple dynamic trait data for each individual in its entirety, and permute the phenotypic data among all individuals. Note that we do not permute the phenotypic data measured at different time points for the same individual.

\section{Application}
We use the proposed functional mapping method to identify QTLs that simultaneously control three dynamic traits of soybeans. These dynamic traits are the time-course biomass of the whole-plant leaf, the whole-plant stem, and the whole-plant root of the soybean. A mapping population composed of 184 recombinant inbred lines (RILs) is derived from the cross of two cultivars, Kefeng No. 1 and Nannong 1138-2. In this RIL population, there are two homozygous genotypes, one containing two Kefeng No. 1 alleles and the other containing two Nannong 1138-2 alleles. A genetic linkage map is constructed with 950 microsatellite markers. The whole-plant leaf biomass, the whole-plant stem biomass, and the whole-plant root biomass were measured for each RIL of the mapping population weekly for eight weeks in a growing season of soybeans.

\begin{figure}
\begin{center}\includegraphics[scale=0.5]{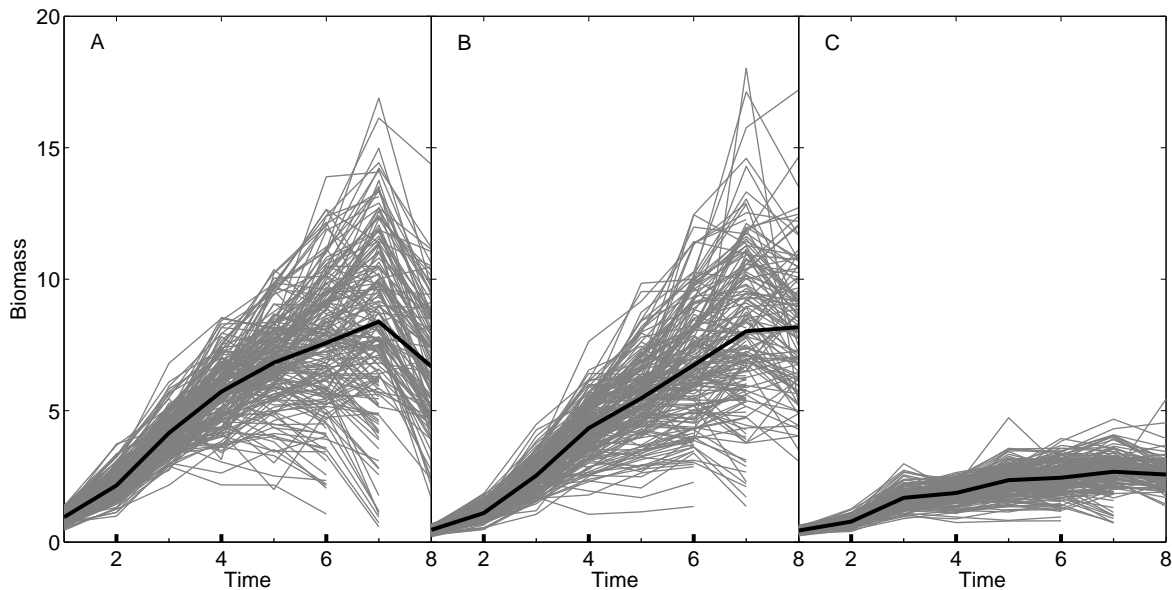}\caption{Dynamic behavior in the whole-plant leaf biomass (A), the whole-plant stem biomass (B), and whole-plant root biomass (C) in a growing season. Each grey line represents the trajectory curve of one of 184 recombinant inbred lines (RILs) and black lines are mean trajectory curves.}\label{Fig:SoybeanData}\end{center}
\end{figure}

Figure \ref{Fig:SoybeanData} illustrates the age-dependent trajectories of the three biomass traits; each of which displays considerable variation. As seen in mature organs with strong cell turnover rates, leaf and root biomass may also undergo reduction as a plant ages. This phenomenon is evident in the age-dependent trajectories of the leaf and root biomasses (see Fig. \ref{Fig:SoybeanData}). Nonparametric regression modeling is employed to fit the dynamic biomass traits in this soybean example, as explained in Section 2.

We scan for possible QTLs by assuming their positions at every 2 cM within a given marker interval in each of 25 linkage groups. For simplicity, we assume a single QTL at a time. For any given position of the QTL, the conditional probability, $\omega_{ij}$, of QTL genotypes is calculated based on the marker genotypes as a function of the recombination fractions between the QTL and markers \shortcite{Wu07}. We then use our proposed profile likelihood method to estimate the mixture model (\ref{Eqn:MixDistribution}). Finally, the LR test statistic (\ref{Eqn:LR}) is calculated at a grid of possible QTL positions. The total computation time for scanning the entire genetic linkage map is around 31 hours by using a MacBook Pro laptop with a 2.5 GHz Intel Core i7 processor and a 16 GB 1600 MHz DDR3 Memory.    

\begin{figure}
\begin{center}\includegraphics[scale=0.55]{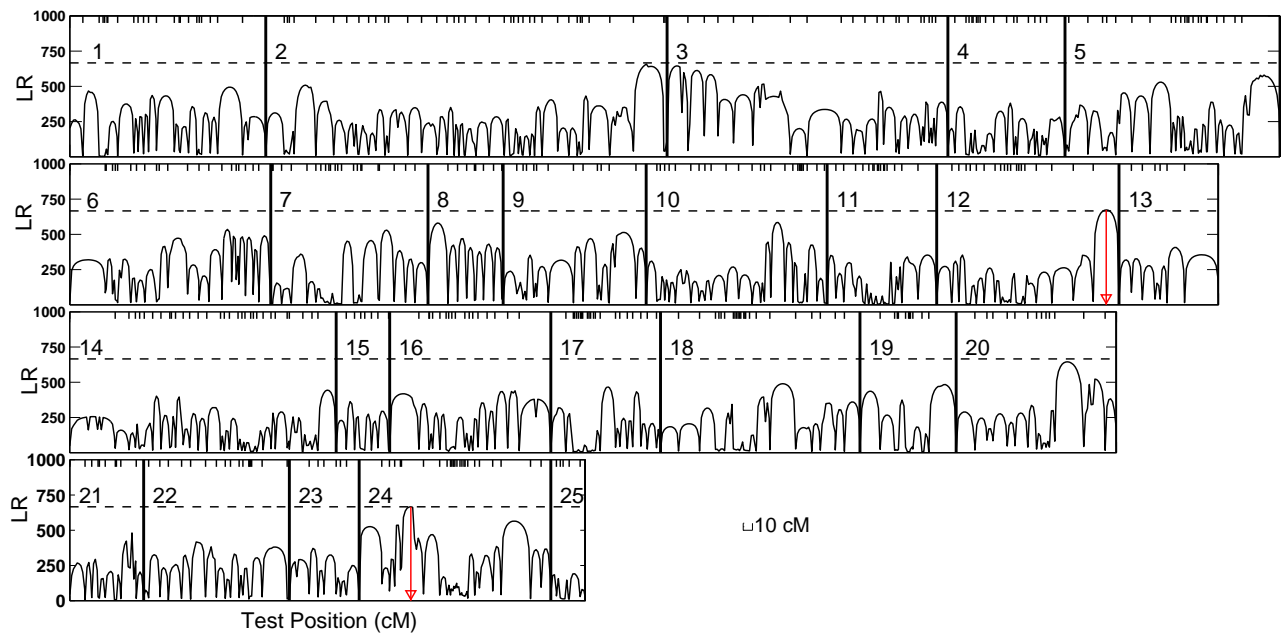}\caption{The log-likelihood ratios (LR) used to test for the existence of QTLs at every 2 cM along the genetic linkage map composed of 950 molecular markers. The number in each panel indicates the 25 linkage groups. The horizontal dashed line indicates the significant threshold for confirming the genome-wide existence of a QTL. Tick marks on the ceilings of each panel represent the positions of molecular markers in each linkage group. The arrowed red lines indicate the locations of the QTLs detected by our method.}\label{Fig:LR}\end{center}
\end{figure}

Figure \ref{Fig:LR} displays the log-likelihood ratio (LR) profile at every 2 cM along the whole genetic linkage map. To determine the significance threshold for confirming the genome-wide existence of a QTL, a permutation test with 100 permutation replicates was conducted. The 95th percentile of the distribution of the maximum LR values obtained from the permutation test is 665.61, which is used as the empirical critical value to declare genome-wide existence of a QTL at the 5\% significance level. The QTL location is estimated by the genomic positions of the peaks of the LR profile that extends beyond the threshold.

\begin{figure}
\begin{center}\includegraphics[scale=0.45]{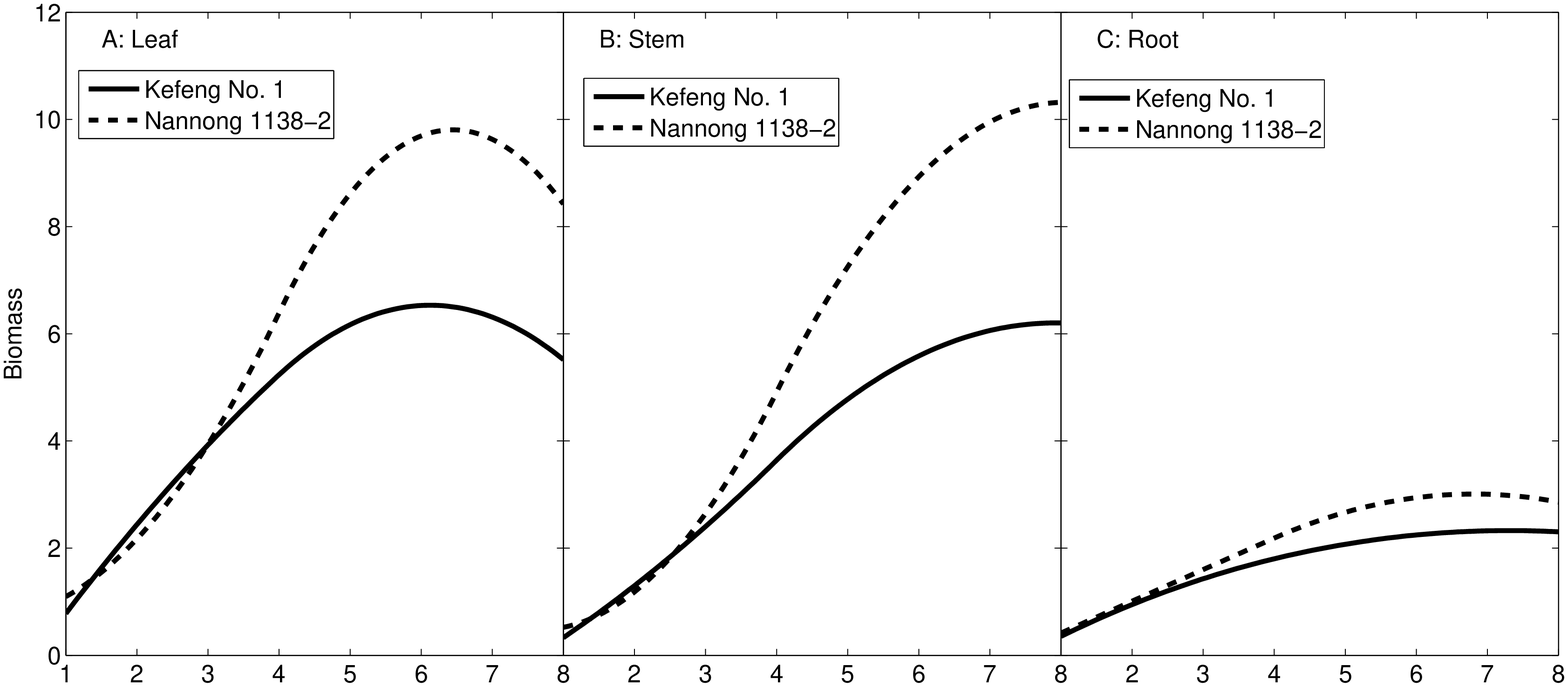}\\\includegraphics[scale=0.45]{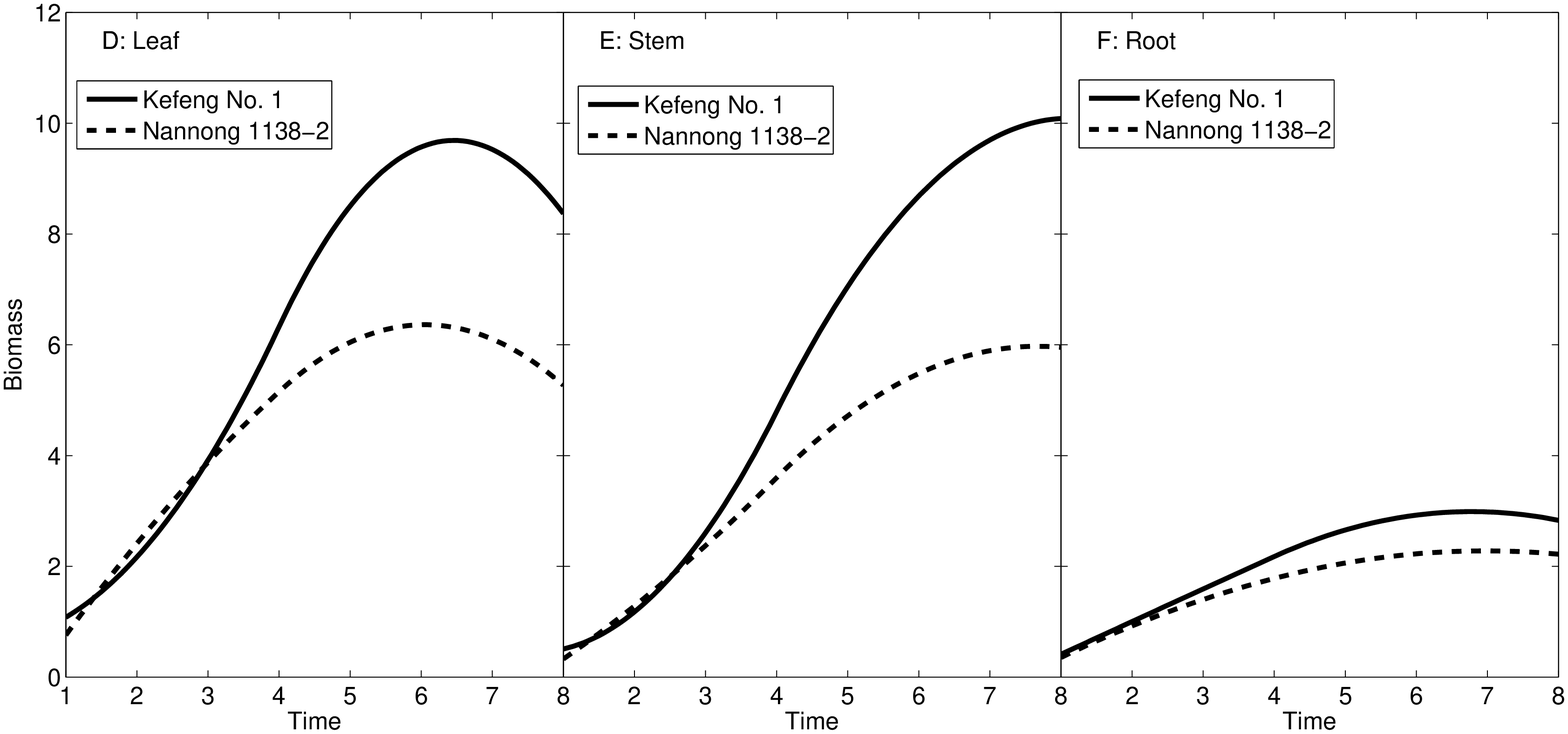}\caption{Estimated mean trajectory curves of whole-plant leaf biomass, stem biomass and whole-plant root biomass for two different genotypes of \texttt{QTL1} detected in linkage group 12 (A, B, C, respectively) and \texttt{QTL2} detected in linkage group 24 (D, E, F, respectively).}\label{Fig:MuHat}\end{center}
\end{figure}

Two significant QTLs are found: one (called \texttt{QTL1}) located between markers GMKF167a and GMKF167b on linkage group 12 and the other (called \texttt{QTL2}) located between markers sat$\_$274 and BE801128 on linkage group 24. The estimated curve parameters for each QTL genotype (${\bf c}_{hj}$) allow us to draw mean biomass trajectories for leaves, the stem and roots, which are displayed in Figure \ref{Fig:MuHat}. At \texttt{QTL1}, the biomasses of the two homozygous genotypes increase in a similar way, but then diverge after the third time of measurement. This suggests that the QTL remains inactive until a particular time point in the growing season. The genotype composed of the parent Nannong 1138-2 alleles displays a much faster rate of increase, especially for leaf and stem biomass, than that of the parent Kefeng No. 1 alleles. The leaf and root biomass of both genotypes decay at a later stage in life. The leaf biomass starts to decay at 6.1 and 6.4 weeks for the genotypes composed of the parent Kefeng No. 1 alleles and Nannong 1138-2 alleles, respectively. A similar pattern is present in the root biomass, which starts to decay at 7.3 and 6.8 weeks, respectively, for the genotypes composed of the parent Kefeng No. 1 alleles and Nannong 1138-2 alleles. Since the leaf and root decay times are slightly different for the two QTL genotypes, this suggests that \texttt{QTL1} may impact the starting time of both leaf biomass and root biomass decay. A similar result is observed for \texttt{QTL2}, except here we see a reversal in the direction of the genetic effects for the two original parents (Fig. \ref{Fig:MuHat}D,E,F). It is interesting to see that the two QTLs detected by our new model have also been observed by our previous model, which integrates allometric scaling through a system of differential equations \shortcite{Wu11}. The distinction between the two models is that our proposed functional mapping method considers the correlation among multiple dynamic traits.

\begin{table}\begin{center}\caption{The estimates and standard errors (SEs) for the standard deviations and correlation coefficients of the biomass of leaves (L), stems (S), and roots (R) for \texttt{QTL1} and \texttt{QTL2}, which are detected on linkage group 12 and 24, respectively. Here, $\sigma$ denotes the standard deviations, and $\rho$ denotes the correlation coefficients. H1 is the alternative hypothesis, which assumes two different mean growth curves for two different genotypes of QTL. H0 is the null hypothesis, which assumes the same mean growth curves for two different genotypes of QTL.}\label{Table:Sigma}\bigskip
\renewcommand{\arraystretch}{1.5}\begin{tabular}{llllllllllllllllll}\hline\hline&&&$\sigma_{LL}$&$\sigma_{SS}$&$\sigma_{RR}$&$\rho_{LS}$&$\rho_{LR}$&$\rho_{SR}$\\\hline
\multirow{4}{*}{\texttt{QTL1}}&\multirow{2}{*}{H1}&Estimate&1.49&1.19&0.41&0.79&0.66&0.65\\\cline{3-9}
&&SE&0.03&0.02&0.01&0.01&0.01&0.02\\\cline{2-9}
&\multirow{2}{*}{H0}&Estimate&1.80&1.67&0.47&0.85&0.74&0.73\\\cline{3-9}
&&SE&0.03&0.03&0.01&0.01&0.01&0.01\\\hline
\multirow{4}{*}{\texttt{QTL2}}&\multirow{2}{*}{H1}&Estimate&1.45&1.20&0.40&0.77&0.67&0.66\\\cline{3-9}
&&SE& 0.03 & 0.02 & 0.01 & 0.01 & 0.01&0.01\\\cline{2-8}\cline{2-9}
&\multirow{2}{*}{H0}&Estimate&1.79&1.66&0.47&0.86&0.76&0.75\\\cline{3-9}
&&SE&0.03&0.03&0.01&0.01&0.01&0.01\\\hline\hline\end{tabular}\end{center}
\end{table}

Table \ref{Table:Sigma} displays the standard deviation and correlation coefficient estimates for the whole-plant leaf biomass, whole-plant stem biomass, and the whole-plant root biomass at \texttt{QTL1} and \texttt{QTL2}. As expected these three traits are positively correlated due to allometric scaling. Strong trait-trait correlations imply the necessity of jointly modeling the three traits. The standard deviations of each trait under the full model (H1, there is a QTL) are 17.2\%, 28.7\%, and 12.8\% smaller than under the reduced model (H0, there is no QTL). Also, comparing the reduced model to the full model, trait-trait correlations tend to decrease under the assumptions of the reduced model. The standard errors of the estimates are obtained by the parametric bootstrap method as follows. For the parameter estimates under H1 and H0, the phenotypic data are generated using the mixture distribution (1) and the multivariate normal distribution, respectively, where the means and the variance-covariance matrices are set to be the same as the estimates from the real data. The standard deviations, $\sigma_{LL}, \sigma_{SS}, \sigma_{RR}$, and correlation coefficients, $\rho_{LS}$, $\rho_{LR}$, $\rho_{SR}$, are then estimated with the profile likelihood method. The above process is replicated 100 times. The sample standard deviation of the 100 replicative estimates for the six parameters are used as the standard errors of the parameter estimates. It shows the standard errors of the estimates are very small, which may indicate that the correlation coefficient estimates are statistically different from zero.

\section{Simulations}
We implement a simulation study to investigate the statistical properties of our functional mapping method. The data are simulated using the same marker information as the twelfth linkage group of the soybean, which is analyzed as a real application in Section 4. This linkage group has 21 markers in total with a length of 196 cM. A QTL, named \texttt{QTL1}, is located at 182.6 cM from the first marker in this group. We assume that three dynamic traits are measured at eight equally-spaced time points, which is also consistent with the real data. The phenotypic data are generated for 184 RILs based on the mixture distribution (\ref{Eqn:MixDistribution}), where means, $\mubold_1(t)$ and $\mubold_2(t)$, and a variance-covariance matrix, $\Sigmabold$, are set to be the same as the estimates from the real data.

For each simulation data set, we scan for possible QTLs by assuming their positions at every 2 cM within a given marker interval in the twelfth linkage group. At each possible location, the conditional probability, $\omega_{ij}$, of QTL genotypes is calculated based on the marker genotypes as a function of the recombination fractions between the QTL and markers \shortcite{Wu07}. Next, we use our proposed profile likelihood method to estimate the mixture model (\ref{Eqn:MixDistribution}), and finally the LR test statistic (\ref{Eqn:LR}) is calculated at any given position of a QTL. We use the permutation test with 100 permutation replicates to obtain the significance threshold for confirming the existence of a QTL. The 95th percentile of the distribution of the maximum LR values is obtained from the permutation test, which is then set to be the empirical critical value for declaring the existence of a QTL at the 5\% significance level. The above simulation procedure is repeated 100 times.

\begin{table}\begin{center}\caption{Means, biases, standard deviations (STDs), and root mean squared errors (RMSEs) of the estimated location of a QTL in 100 simulation replicates using the functional mapping method with or without considering the correlation of multiple dynamic traits. }\label{Table:SimuQTLlocation}\vspace{1cm}
\renewcommand{\arraystretch}{1.5}\begin{tabular}{rrrrrrrrrrr}\hline\hline
&True&Mean&Bias&STD&RMSE&Confidence Interval\\\hline 
Correlated Model&182.6&182.4&0.2&1.9&1.9&[178.7,186.2]\\\hline 
Uncorrelated Model&182.6&180.6&2.0&18.3&18.4&[144.7, 216.6]\\\hline
\end{tabular}\end{center}
\end{table}

We compare our functional mapping method, which accounts for correlations, with the traditional method, which does not consider the correlations between multiple dynamic traits. Table \ref{Table:SimuQTLlocation} summarizes the estimated QTL locations in 100 simulation replicates for each of these two methods. When considering the correlation between multiple dynamic traits, we see that the biases, standard deviations (STDs) and root mean squared errors (RMSEs) of the estimated QTL locations are reasonably small, which indicates that our functional mapping method provides an accurate estimate of the QTL location for this sample size. Alternatively, if we do not consider the correlation between multiple dynamic traits, the estimate of the QTL position is very biased. The standard deviation and root mean squared error of the estimated QTL locations also substantially increase by not considering the correlation between multiple dynamic traits. Notably, the RMSE of the estimated QTL locations is decreased by 89.7\% by considering the correlation between multiple dynamic traits.

\begin{figure}
\begin{center}
\includegraphics[scale=0.4]{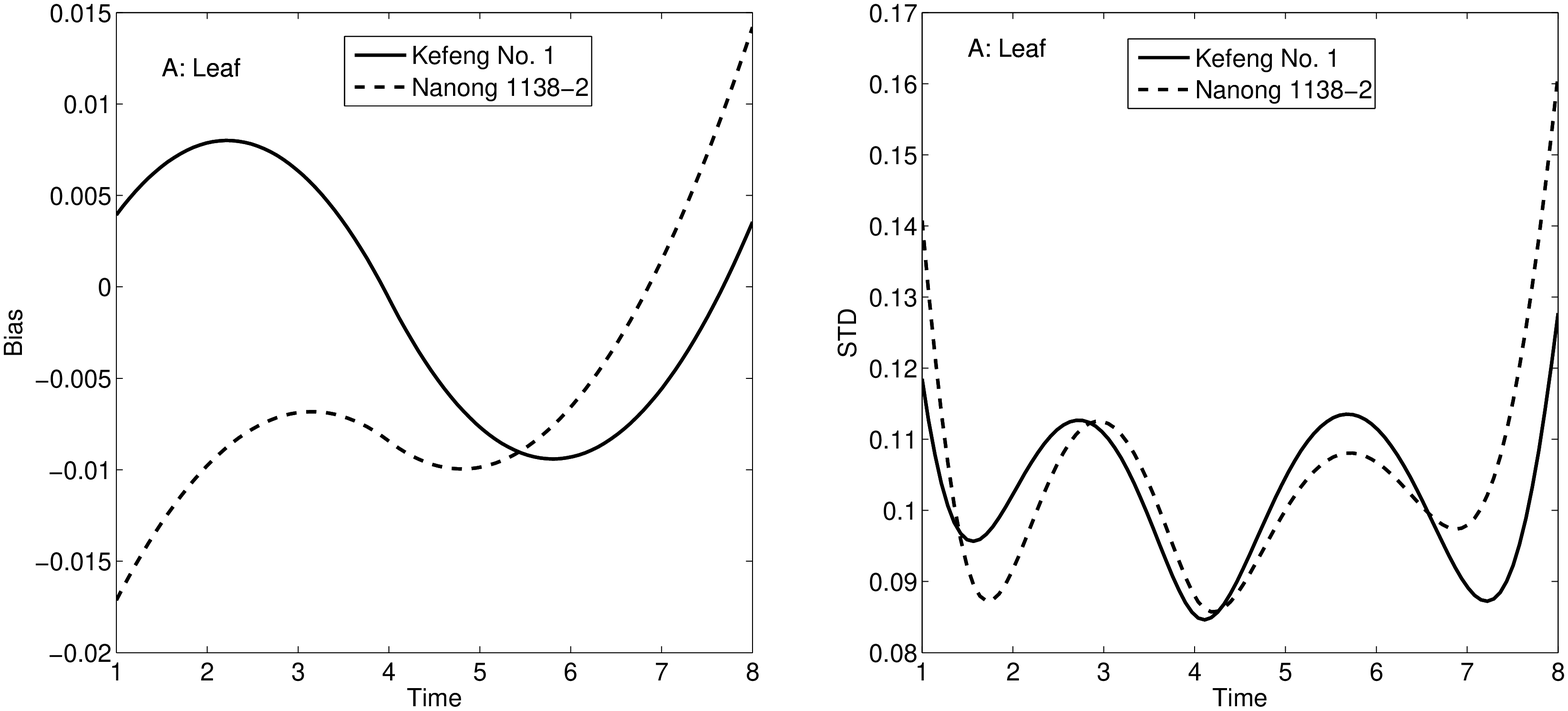}
\includegraphics[scale=0.4]{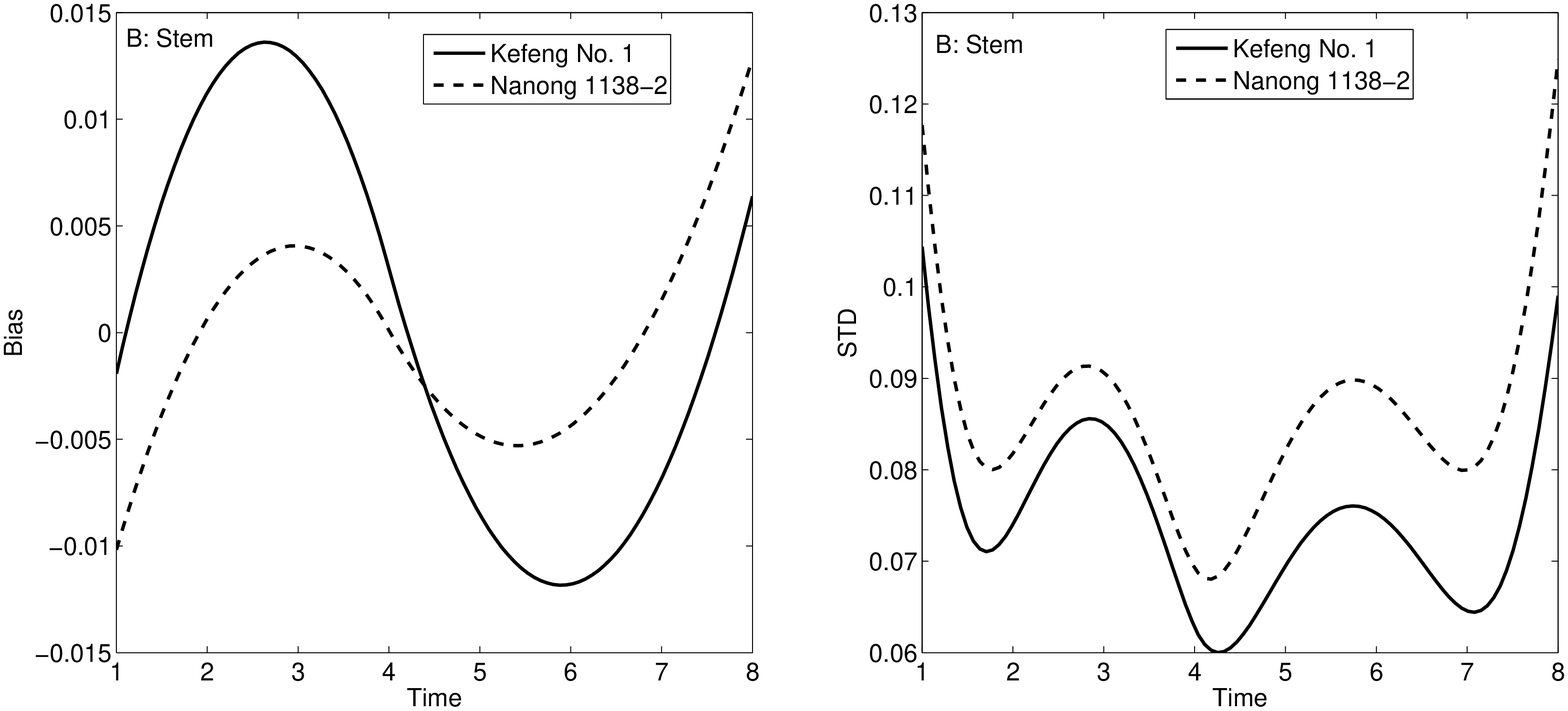}
\includegraphics[scale=0.4]{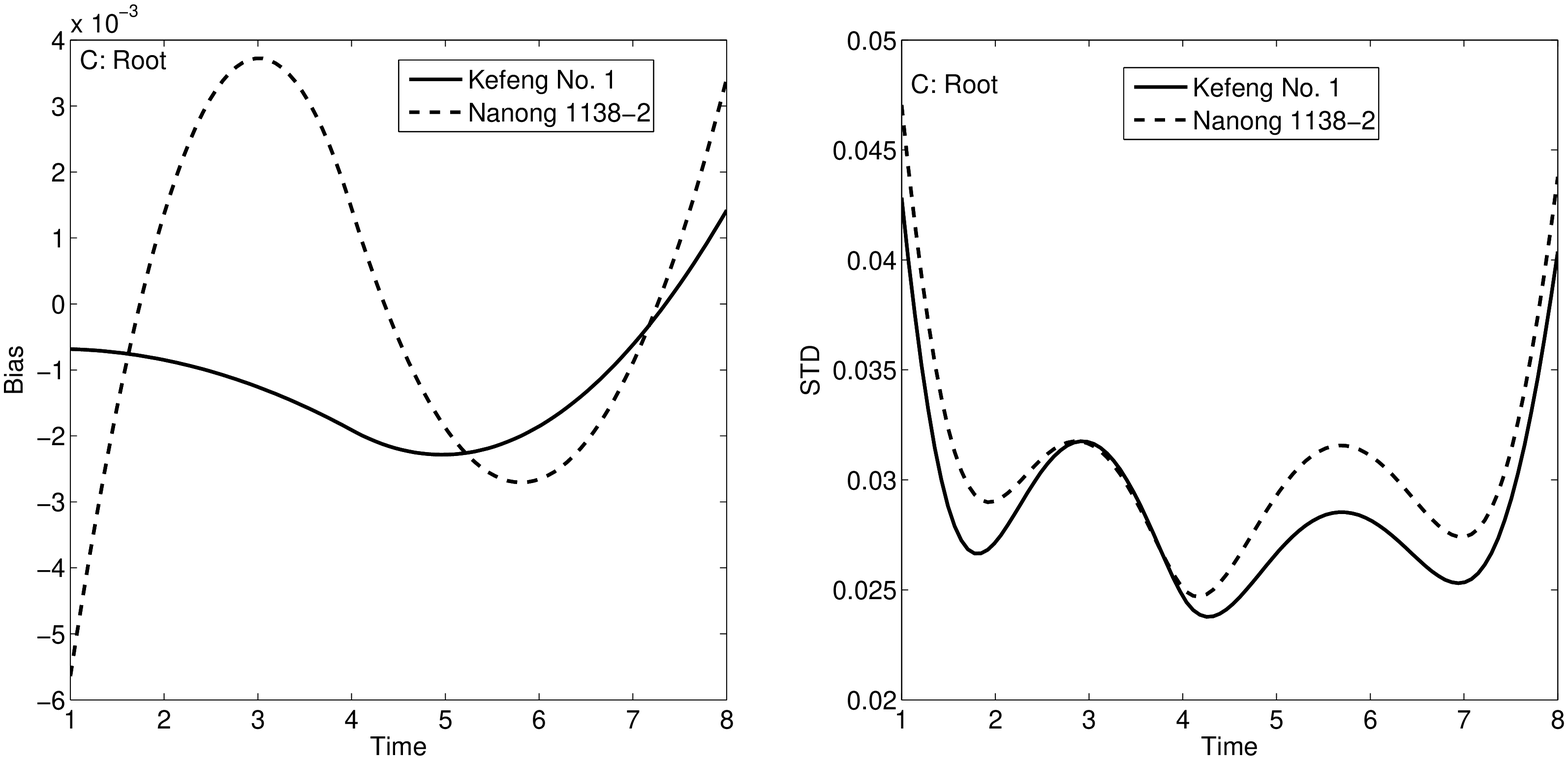}
\caption{The point-wise biases and standard deviations (STDs) of the estimated mean trajectory curves of whole-plant leaf biomass, stem biomass and whole-plant root biomass (marked by A, B, C, respectively) for two different genotypes (plotted in solid and dashed lines, respectively,) of \texttt{QTL} located at 182 cM from the first marker in linkage group 12 in our simulation study.}\label{Fig:BiasMeanCurve}\end{center}
\end{figure}

Figure \ref{Fig:BiasMeanCurve} displays the point-wise biases and standard deviations (STDs) of the estimated mean trajectory curves of the whole-plant leaf biomass, the whole-plant stem biomass, and the whole-plant root biomass for the two different genotypes of the \texttt{QTL} located at 182 cM from the first marker in the twelfth linkage group. It can be seen that the biases of the estimated mean trajectory curves are negligible for the whole-plant leaf, stem, and root.

A power study is implemented to evaluate the power of the proposed likelihood ratio test to determine the existence of a QTL. Assuming a QTL is located at 182.6 cM from the first marker in the twelfth linkage group of the soybean, the phenotypic data are generated for 184 RILs from the mixture distribution with the true parameter values same to the estimates from the real data. When a QTL exists, i.e., the alternative hypothesis is true, we detect a QTL in all simulation replicates. So the power of the proposed test is 100\%.

\section{Discussion}
Genetic mapping techniques have developed to a point where it is crucial to implement systematic modeling of phenotypic information to better understand the developmental mechanisms of biological processes. This requires an integral merger of genetic mapping with developmental principles through robust statistical and mathematical models. Over the last decade, a series of statistical models, packed as functional mapping, have been proposed to map quantitative trait loci (QTLs) that mediate the developmental pattern and form of phenotypic traits, facilitating our insight into the causal interplay between genes and development \shortcite{Ma02,Zhao05,Lietal06,WuLin06,Lin06,Yang07,LiWu10,He10,Yang09,Wu11}. These models show their unique power under specific circumstances.

In this article, we develop an innovative version of functional mapping by implementing a multivariate mixture model. The biological merit of this innovation is two-fold: (1) it is highly flexible to fit any form of trajectory curves and (2) it allows multiple dynamic traits to be analyzed simultaneously, providing a general way to test for pleiotropic control of QTLs. We have for the first time both derived a statistical method for estimating the multivariate mixture model and studied its statistical properties.

Perhaps, the most significant part of this study lies in its scientific validation and application to a real data set for QTL mapping in soybeans. The two significant QTLs detected by our new model have very intuitive interpretations that agree with developmental principles of trait formation and progression. It is impossible to obtain such an in-depth understanding of trait control by traditional QTL mapping approaches based on static traits. The new model allows numerous versatile tests for when and how a QTL exerts its effect on trait development. If a trait, such as leaf biomass or root biomass, experiences growth and senescence stages, the new model is able to test whether the detected QTLs determine the timing of a trait's developmental transitions. 

A linear or semiparametric mixed model is a possible alternative to test for the effects of genetic markers on the multivariate traits \shortcite{thiebaut2002bivariate,sithole2007bivariate,ghosh2010semiparametric,das2014semiparametric}. However, this framework is based on the assumption that QTLs controlling multivariate traits are observed directly, and are included in observed genetic markers. This assumption is beyond the scope of this manuscript. 

\section*{Acknowledgements}This research is supported by NSF/IOS-0923975 to R. Wu and discovery grants of the Natural Sciences and Engineering Research Council of Canada (NSERC) to J. Cao and L. Wang.

\bibliographystyle{chicago}
\bibliography{References}
\end{document}